# Inverse problem for Ising connection matrix with long-range interaction

L.B. Litinskii and B.V. Kryzhanovsky

Center of Optical Neural Technologies, Scientific Research Institute for System Analysis RAS, Moscow, Russia
Nakhimov ave, 36-1, Moscow, 117218, Russia

E-mail: litin@mail.ru, kryzhanov@mail.ru

**Abstract**

In the present paper, we examine Ising systems on $d$-dimensional hypercube lattices and solve an inverse problem where we have to determine interaction constants of an Ising connection matrix when we know a spectrum of it eigenvalues. In addition, we define restrictions allowing a random number sequence to be a connection matrix spectrum. We use the previously obtained analytical expressions for the eigenvalues of Ising connection matrices accounting for an arbitrary long-range interaction and supposing periodic boundary conditions.

## 1. Introduction

In papers [1–3], we calculated eigenvalues of Ising connection matrices defined on $d$-dimensional hypercube lattices ($d = 1, 2, 3...$). To provide the translation invariance we imposed periodic boundary conditions. In our calculations, we accounted for interactions not only with the nearest spins but with distant spins too. In papers [1, 2] we analyzed isotropic interactions. The general case of anisotropic interactions we discussed in [3]. We succeeded to obtain analytical expressions for the eigenvalues of the above-described Ising connection matrices. For the $d$-dimensional system, the eigenvalues are polynomials of the degree $d$ in the eigenvalues for the one-dimensional system with long-range interaction (see [2, 3]). The coefficients of these polynomials are the constants of interaction between spins.

In the present paper, we solve an inverse problem formulated as follows. Suppose we know the spectrum of an Ising connection matrix and we have to answer two questions. Firstly, is it possible to restore the interaction constants that define the connection matrix whose spectrum matches the given one? Secondly, when a sequence of random numbers may be the spectrum of some connection matrix? In Section 2, we obtain the answers to these questions for the one-dimensional Ising system. Then, we use the obtained results to analyze the two- and three-dimensional systems in Sections 3 and 4, respectively. Discussion and conclusions are in Section 5. In this section, we also discuss the possibilities to use the eigenvalues of the Ising connection matrices when calculating the partition functions.

There is extensive literature on *the inverse Ising problem*; see, for example, a rather full review published in [7]. When solving the inverse Ising problems the authors examine how with the aid the statistical inference method they can estimate the parameters of the Ising system - interaction constants and external magnetic fields - when they know empirical characteristics of a large number of random spin configurations. We would like to emphasize that although as in the papers cited in [7] we also restoring the parameters of the Ising systems, the setting of the problem and the method of its solution differ significantly. In our approach we inverse the exact formulas that express the connection matrix eigenvalues in terms of its matrix elements. However, when using the statistical inference method the input data are the observables such as magnetizations, correlations, and so on. The solution tools are also different. They are the Boltzmann equilibrium distribution, the principle of the maximal likelihood, the Bayes theorem and so on.

## 2. One-dimensional Ising model

**1)** A one-dimensional Ising system is a linear chain of $L$ interacting spins. To provide a translation invariance, let us close the chain in a ring. Then the last spin is also the nearest neighbor of the first spin. This means that each spin has two (on the left and right) nearest neighbors, two next nearest neighbors (the distance to which is twice as large), two next-next nearest neighbors, and so on. To be specific, we suppose that $L$ is odd: $L = 2l + 1$. Consequently, each spin has $l$ pairs of the neighbors. Since we have in mind to discuss multidimensional lattices, we use the term "coordination spheres" to describe these pairs: first coordination sphere, second coordination sphere … $l$-th coordination sphere. In the beginning of the next Section, we will give a general definition of the coordination spheres.

By $\mathbf{J}(k)$ we denote a connection matrix that defines the interaction of each spin *only* with the spins from the $k$-th coordination sphere. For example, it is easy to see that the matrices $\mathbf{J}(1)$ and $\mathbf{J}(2)$ have the form

$$\mathbf{J}(1) = \begin{pmatrix} 0 & 1 & 0 & 0 & . & . & 1 \\ 1 & 0 & 1 & 0 & . & . & 0 \\ 0 & 1 & 0 & 1 & . & . & 0 \\ . & . & . & . & . & . & . \\ 0 & 0 & . & . & 1 & 0 & 1 \\ 1 & 0 & 0 & . & . & 1 & 0 \end{pmatrix}, \quad \mathbf{J}(2) = \begin{pmatrix} 0 & 0 & 1 & 0 & . & 1 & 0 \\ 0 & 0 & 0 & 1 & . & . & 1 \\ 1 & 0 & 0 & 0 & . & . & 0 \\ . & . & . & . & . & . & 1 \\ 1 & 0 & . & 1 & 0 & 0 & 0 \\ 0 & 1 & 0 & . & 1 & 0 & 0 \end{pmatrix}.$$

$\mathbf{J}(k)$ is a symmetric matrix with the ones at the $k$-th and $(L-k)$-th diagonals that are parallel to the main diagonal. We use the set of matrices $\{\mathbf{J}(k)\}_1^l$ to write down the Ising connection matrix $\mathbf{A}_0$ that accounts for interactions with spins belonging to all the coordination spheres. Let $w_k$ be a constant of interaction with spins from the $k$-th coordination sphere. Then

$$\mathbf{A}_0 = w_1 \cdot \mathbf{J}(1) + w_2 \cdot \mathbf{J}(2) + ... + w_l \cdot \mathbf{J}(l). \tag{1}$$

When there is no interaction with the spins from the $k$-th coordination sphere, the corresponding constant $w_k$ in Eq. (1) is equal to zero.

2) The matrices $\mathbf{J}(k)$ are circulants: each next row of such a matrix is obtained by a cyclic shift of the previous row one position to the right. All the circulants have the same set of the eigenvectors that may have complex coordinates [5, 6]. In the general case, the eigenvalues of the circulant matrices can also be complex. However, since in our problem the matrices $\mathbf{J}(k)$ are symmetric, their eigenvalues are real. By $\{\lambda_\alpha(k)\}_{\alpha=1}^L$ we denote the eigenvalues of these matrices. It can be shown that [2, 3]

$$\lambda_\alpha(k) = 2 \cdot \cos(k\varphi_\alpha), \text{ where } \varphi_\alpha = \frac{2\pi}{L}(\alpha-1), \alpha = 1, 2, ..., L, \text{ and } k = 1, 3, ..., l.$$

The first eigenvalue of each matrix $\mathbf{J}(k)$ is equal to 2, and other eigenvalues are twice degenerate:

$$\forall k = 1, 2, ..., l: \ \lambda_1(k) = 2, \text{ and } \lambda_\alpha(k) = \lambda_{L+2-\alpha}(k), \text{ where } \alpha = 2, 3, ..., l+1. \tag{2}$$

Consequently, for each $k$ (if we do not take into account the first eigenvalue), the sequence $\lambda_2(k), \lambda_3(k), ..\lambda_{l+1}(k), \lambda_{l+2}(k), .., \lambda_{L-1}(k), \lambda_L(k)$ is mirror-symmetrical about its middle (see Fig. 1). In what follows, we repeatedly use this symmetry property.

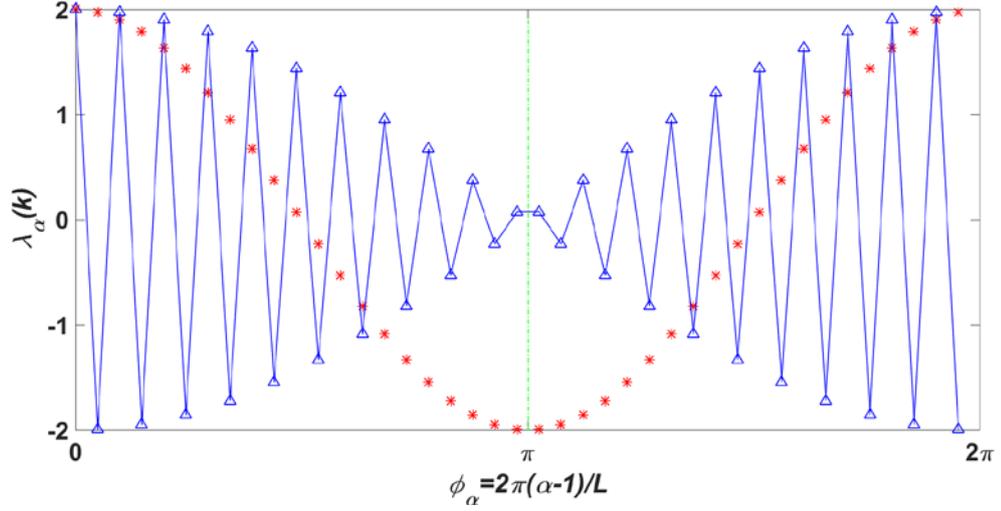

Fig. 1. Eigenvalues $\{\lambda_\alpha(k)\}$, $\alpha = 1,2,...,L = 41$ of matrices $\mathbf{J}(k)$ when $k = 1$ ($*$) and $k = 20$ ($\Delta$).
Vertical line in the middle shows explicitly mirror symmetry of graphs.

The eigenvector $\mathbf{f}^{(1)}$ with equal coordinates corresponds to the first eigenvalue $\lambda_1(k) = 2$:

$$\mathbf{J}(k) \cdot \mathbf{f}^{(1)} = \lambda_1(k) \cdot \mathbf{f}^{(1)}, \text{ where } \mathbf{f}^{(1)} = \begin{pmatrix} 1 \\ 1 \\ \vdots \\ 1 \end{pmatrix} / \sqrt{L}.$$

We can choose the two eigenvectors $\mathbf{f}^{(\alpha)}$ and $\mathbf{f}^{(L+2-\alpha)}$ corresponding to a degenerate eigenvalue $\lambda_\alpha(k) = \lambda_{L+2-\alpha}(k)$ to be real. It is convenient to write them as

$$\mathbf{f}^{(\alpha)} = \begin{pmatrix} f_1^{(\alpha)} \\ f_2^{(\alpha)} \\ \vdots \\ f_L^{(\alpha)} \end{pmatrix}, \quad f_j^{(\alpha)} = \sqrt{\frac{2}{L}} \cos\left((j-1)\varphi_\alpha\right)$$

$$\mathbf{f}^{(L+2-\alpha)} = \begin{pmatrix} f_1^{(L+2-\alpha)} \\ f_2^{(L+2-\alpha)} \\ \vdots \\ f_L^{(L+2-\alpha)} \end{pmatrix}, \quad f_j^{(L+2-\alpha)} = \sqrt{\frac{2}{L}} \sin\left((j-1)\varphi_\alpha\right)$$

, where $\alpha = 2,3,...,l+1$. (3)

Since the eigenvectors of all the matrices $\mathbf{J}(k)$ are the same, it is easy to write down the eigenvalues of the connection matrix (1)

$$\lambda_\alpha(\mathbf{A}_0) = w_1 \cdot \lambda_\alpha(1) + w_2 \cdot \lambda_\alpha(2) + ... + w_l \cdot \lambda_\alpha(l), \quad \alpha = 1,2,...,L. \quad (4)$$

The expression (4) is a generalization of the formula obtained previously in [4].

The spectrum of the eigenvalues of the connection matrix $\mathbf{A}_0$ cannot be a set of arbitrary numbers. It has a structure defined by the properties of the summands in Eq. (4). First, because the equalities (2) hold for each $k$, the spectrum of the eigenvalues $\{\lambda_\alpha(\mathbf{A}_0)\}_{\alpha=1}^L$ has to be mirror-symmetrical about its middle (without accounting for the first eigenvalue). Then we have the equalities

$$\lambda_\alpha(\mathbf{A}_0) = \lambda_{L+2-\alpha}(\mathbf{A}_0), \quad \alpha = 2,3,...,l+1. \quad (5)$$

Second, due to the zero-valued elements at the diagonals of all the matrices $\mathbf{J}(k)$ the sum of the eigenvalues of the matrix $\mathbf{A}_0$ has to be equal to zero. This means that

$$\lambda_1(\mathbf{A}_0) = -2\sum_{\alpha=2}^{l+1} \lambda_\alpha(\mathbf{A}_0). \tag{6}$$

Consequently, only $l$ numbers $\lambda_2(\mathbf{A}_0)$, $\lambda_3(\mathbf{A}_0)$, ... $\lambda_{l+1}(\mathbf{A}_0)$ of the set (4) can be arbitrary; the other eigenvalues are expressed through these numbers with the aid of the equalities (5) and (6).

**3)** Let us analyze the inverse problem. Suppose we know a spectrum $\{\tilde{\lambda}_\alpha\}_{\alpha=1}^L$ of a connection matrix of a one-dimensional Ising system (for example, obtained experimentally). Of course, the sequence $\{\tilde{\lambda}_\alpha\}_{\alpha=1}^L$ satisfies the equalities (5) and (6). What are the connections $w_k$ between the spins that provide this spectrum?

To determine the unknowns $w_k$, we have to solve the system (4) with the known left-hand side:

$$\tilde{\lambda}_\alpha = w_1 \cdot \lambda_\alpha(1) + w_2 \cdot \lambda_\alpha(2) + ... + w_l \cdot \lambda_\alpha(l), \quad \alpha = 1, 2, ..., L. \tag{7}$$

We can obtain the answer in an explicit form. Let us generate an $L$-dimensional vector $\tilde{\boldsymbol{\Lambda}}$ whose coordinates are the eigenvalues of the experimental spectrum $\{\tilde{\lambda}_\alpha\}_{\alpha=1}^L$. We also generate $L$-dimensional vectors $\boldsymbol{\Lambda}(k)$ whose coordinates are the eigenvalues of the matrices $\mathbf{J}(k)$:

$$\tilde{\boldsymbol{\Lambda}} = \begin{pmatrix} \tilde{\lambda}_1 \\ \tilde{\lambda}_2 \\ \vdots \\ \tilde{\lambda}_{l+1} \\ \tilde{\lambda}_{l+2} \\ \vdots \\ \tilde{\lambda}_L \end{pmatrix}, \quad \boldsymbol{\Lambda}(k) = \begin{pmatrix} \lambda_1(k) \\ \lambda_2(k) \\ \vdots \\ \lambda_{l+1}(k) \\ \lambda_{l+2}(k) \\ \vdots \\ \lambda_L(k) \end{pmatrix}, \quad k = 1, 2, ..., l. \tag{8}$$

Then we can rewrite the system of equations (7) in the vector form:

$$\tilde{\boldsymbol{\Lambda}} = w_1 \cdot \boldsymbol{\Lambda}(1) + w_2 \cdot \boldsymbol{\Lambda}(2) + ... + w_l \cdot \boldsymbol{\Lambda}(l).$$

It is evident that the vectors $\boldsymbol{\Lambda}(k)$ and the eigenvectors $\mathbf{f}^{(k+1)}$ are collinear: $\boldsymbol{\Lambda}(k) \sim \mathbf{f}^{(k+1)}$, $k = 1, 2, ..., l$. Consequently, we can calculate the weights $w_k$ as scalar products of the vectors $\tilde{\boldsymbol{\Lambda}}$ and $\boldsymbol{\Lambda}(k)$:

$$w_k = \frac{(\tilde{\boldsymbol{\Lambda}}, \boldsymbol{\Lambda}(k))}{\sqrt{2L}}, \quad k = 1, 2, ..., l. \tag{9}$$

By doing that, we solve the inverse problem in the one-dimensional case.

**3. Two-dimensional Ising model**

**1)** In this case, the spins are in the nods of a square lattice of the size $L \times L$. As previously, we set $L = 2l + 1$ and assume periodic boundary conditions. Then each spin has $l$ pairs of neighbors along both the horizontal and the vertical axes. In addition, there are neighbors that are not on the same horizontal or vertical axes as the given spin.

The set of spins equally interacting with the given spin belongs to the same coordination sphere. In the case of an isotropic interaction, the coordination spheres consist of spins equally distant from the given spin. Then we can enumerate the coordination spheres in the ascending order of distances to the given spin. In the anisotropic case the interaction constants but not the distances define the spins belonging to the given coordination sphere.

When analyzing multidimensional Ising systems, we first have to distribute spins between the coordination spheres. This step is simple in the one-dimensional case: the pair of spins that are equidistant from the given spin belongs to the same coordination sphere. In the case of two-dimensional lattice, to describe the interaction between the spins spaced by $m$ steps along the vertical axis and by $k$ steps along the horizontal axis we introduce the interaction constant $w(m,k)$; the values of $m$ and $k$ change independently from 0 to $l$. If the interaction is anisotropic $w(m,k) \neq w(k,m)$; in the isotropic case $w(m,k) \equiv w(k,m)$. The difference between the coordination spheres in the isotropic and anisotropic cases influences symmetry properties of the spectrum.

Let us make a few necessary comments. Since there is no a self-action in the system, we always have $w(0,0) = 0$. It is convenient to introduce a unit $(L \times L)$-dimensional matrix $\mathbf{J}(0) = \mathbf{diag}(1,1,..,1)$. This matrix completes the set of matrices $\{\mathbf{J}(k)\}_{k=1}^{l}$. All the eigenvalues of the matrix $\mathbf{J}(0)$ are equal to one. With the aid of these eigenvalues, we define the $L$-dimensional vector

$$\mathbf{\Lambda}(0) = \begin{pmatrix} 1 \\ 1 \\ \vdots \\ 1 \end{pmatrix},$$

which completes the set (8) of the vectors $\mathbf{\Lambda}(k)$: $\{\mathbf{\Lambda}(k)\}_{k=0}^{l}$.

In subsection 2), we solve the inverse problem in the case of anisotropic interaction. The isotropic interaction is a subject of the last subsection of this Section.

**2)** In paper [3] we showed that a $(L^2 \times L^2)$-dimensional matrix $\mathbf{B}_0$ that described the interactions $\{w(m,k)\}_{k,m=0}^{l}$ between spins had a block-circulant form and it eigenvectors were the pairwise Kronecker products of the eigenvectors $\mathbf{f}^{(\alpha)}$ defined by Eq. (3). Exactly as in the one-dimensional case, the set of the eigenvectors of the matrix $\mathbf{B}_0$ does not depend on the interaction constants and the eigenvalues of this matrix obtained in [3] are

$$\mu_{\alpha\beta} = \sum_{m=0}^{l} \sum_{k=0}^{l} w(m,k) \cdot \lambda_\alpha(m) \cdot \lambda_\beta(k), \ \alpha, \beta = 1, 2, ..., L. \tag{10}$$

Let us write Eq. (10) in the vector form using the above-introduced $L$-dimensional vectors $\mathbf{\Lambda}(k)$ (see Eq. (8)). With the aid of these vectors we generate $L^2$-dimensional vectors $\mathbf{\Lambda}(m,k)$ that are the Kronecker products of the vectors $\mathbf{\Lambda}(m)$ and $\mathbf{\Lambda}(k)$:

$$\mathbf{\Lambda}(m,k) = \mathbf{\Lambda}(m) \otimes \mathbf{\Lambda}(k) = \begin{pmatrix} \lambda_1(m)\mathbf{\Lambda}(k) \\ \lambda_2(m)\mathbf{\Lambda}(k) \\ \vdots \\ \lambda_{l+1}(m)\mathbf{\Lambda}(k) \\ \lambda_{l+2}(m)\mathbf{\Lambda}(k) \\ \vdots \\ \lambda_L(m)\mathbf{\Lambda}(k) \end{pmatrix}, \ m,k = 0,1,..,l. \tag{11}$$

The vectors $\mathbf{\Lambda}(m,k)$ are mutually orthogonal. Let us define an $L^2$-dimensional vector $\mathbf{M}$ whose coordinates are the eigenvalues $\mu_{\alpha\beta}$ defined by Eq. (10):

$$\mathbf{M} = (\mu_{11}..\mu_{1L}, \mu_{21}..\mu_{2L}, ..., \mu_{l+11}..\mu_{l+1L}, \mu_{l+21}..\mu_{l+2L}, ..., \mu_{L1}..\mu_{LL})^+. \tag{12}$$

Now we can rewrite the set of equalities (10) in the vector form

$$\mathbf{M} = \sum_{m=0}^{l}\sum_{k=0}^{l} w(m,k)\cdot \mathbf{\Lambda}(m,k) = w(0,1)\cdot \mathbf{\Lambda}(0,1) + ... + w(0,l)\cdot \mathbf{\Lambda}(0,l) + \\ + w(1,0)\cdot \mathbf{\Lambda}(1,0) + ... + w(l,0)\cdot \mathbf{\Lambda}(l,0) + ...... + w(l,0)\cdot \mathbf{\Lambda}(l,0) + ... + w(l,l)\cdot \mathbf{\Lambda}(l,l).$$ (13)

Since $w(0,0) = 0$, in this equation the term $w(0,0)\cdot \mathbf{\Lambda}(0,0)$ is absent.

Equation (13) allows us to solve easily the two-dimensional inverse problem. Namely, we have to determine the interaction constants $\{w(m,k)\}$ that provide a known eigenvalues spectrum $\{\tilde{\mu}_{\alpha\beta}\}_{\alpha,\beta=1}^{L}$. For example, it might be an experimental spectrum.

Let us write an $L^2$-dimensional column vector $\tilde{\mathbf{M}}$ of the form (12) using the "experimental" spectrum components $\{\tilde{\mu}_{\alpha\beta}\}_{\alpha,\beta=1}^{L}$ and let us take into account the mutual orthogonality of the vectors $\mathbf{\Lambda}(m,k)$ (11). Then the desired interaction constants are the scalar products of the $L^2$-dimensional vectors:

$$w(m,k) = \frac{\left(\tilde{\mathbf{M}}, \mathbf{\Lambda}(m,k)\right)}{\|\mathbf{\Lambda}(m,k)\|}, \quad m,k = 0,1,...l.$$ (14)

Now let us discuss another question. In the same way as in the one-dimensional problem, not any sequence of the numbers $\{\mu_{\alpha\beta}\}_{\alpha,\beta=1}^{L}$ can be a spectrum of a connection matrix: the symmetry properties of the $L^2$-dimensional vectors $\mathbf{\Lambda}(m,k)$ impose rather severe restrictions on the values of these numbers.

Firstly, from Eq. (13) it follows that the sum of the numbers $\mu_{\alpha\beta}$ has to be equal to zero:

$$\sum_{\alpha=1}^{L}\sum_{\beta=1}^{L} \mu_{\alpha\beta} = 0.$$ (15)

Secondly, in the one-dimensional problem the set of the eigenvalues (excluding the first eigenvalue) is mirror-symmetrical about its middle for each $m = 0,1,..,l$:

$$\lambda_i(m) = \lambda_{L+2-i}(m), \text{ where } i = 2,3,...,l+1.$$

From Eq. (11), which defines the $L^2$-dimensional vectors $\mathbf{\Lambda}(m,k)$ as the products of the eigenvalues $\lambda_i(m)$ by the vectors $\mathbf{\Lambda}(k)$, it is evident that their last $l\cdot L$ coordinates copy the preceding $l\cdot L$ ones. Consequently, the same has to be true for the sequence of the numbers $\{\mu_{\alpha\beta}\}_{\alpha,\beta=1}^{L}$. Then it is necessary that the numbers that constituting the spectrum satisfy the equalities

$$\mu_{\alpha\beta} = \mu_{L+2-\alpha\,\beta}, \text{ where } \alpha = 2,...,l+1, \beta = 1,L.$$

In other words, the last $l\cdot L$ terms of the sequence of the numbers $\mu_{\alpha\beta}$ are not free parameters.

Thirdly, since the last $l$ coordinates of each $L$-dimensional vector $\mathbf{\Lambda}(k)$ are a mirror image of the preceding $l$ coordinates, not all the first $(l+1)\cdot L$ coordinates of any vector $\mathbf{\Lambda}(m,k)$ are different. Consequently, the same has to be true for the given sequence $\{\mu_{\alpha\beta}\}_{\alpha,\beta=1}^{L}$: the last $l$ terms of the first group of it $L$ terms have to be a mirror image of the preceding $l$ terms; the last $l$ terms of the second group of it $L$ terms have to be a mirror image of the preceding $l$ terms, and so on. Finally, for the last $(l+1)$-th group consisting of $L$ terms of the sequence $\{\mu_{l+1,\beta}\}_{\beta=1}^{L}$ the equalities $\mu_{l+1,i} = \mu_{l+1,L+2-i}$ ($i = 2,3,...,l+1$) have to be fulfilled. This means that by symmetry reasons only $(l+1)^2$ numbers

$$\{\mu_{1\beta}\}_{\beta=1}^{l+1}, \{\mu_{2\beta}\}_{\beta=1}^{l+1}, \{\mu_{3\beta}\}_{\beta=1}^{l+1},...,\{\mu_{l+1\beta}\}_{\beta=1}^{l+1}$$ (16)

of the sequence $\{\mu_{\alpha\beta}\}_{\alpha,\beta=1}^{L}$ may be independent parameters.

We can rewrite Eq. (15) using only the terms of the sequence (16):

$$\mu_{11} + 2\left(\sum_{\beta=2}^{l+1}\mu_{1\beta} + \sum_{\alpha=2}^{l+1}\mu_{\alpha 1}\right) + 4\sum_{\alpha=2}^{l+1}\sum_{\beta=2}^{l+1}\mu_{\alpha\beta} = 0. \qquad (17)$$

This equation allows us to express $\mu_{11}$, through the other $l(l+2)$ independent numbers $\mu_{\alpha\beta}$ from the sequence (16). Consequently, the number of the independent values $\mu_{\alpha\beta}$ equals exactly to the number of the orthogonal vectors $\Lambda(n,m)$ taking part in the expansion (13).

**3)** Finally, let us discuss briefly a two-dimensional Ising system with an isotropic interaction. Evidently, we again can use the equations (13), (14), and (17), however, now the number of various interaction constants $w(m,k)$ is not $l(l+2)$ but much less: their number is equal to $l(l+3)/2$. This means that the same has to be the number of independent terms in the given sequence $\{\mu_{\alpha\beta}\}_{\alpha,\beta=1}^{L}$ that represents the spectrum of an isotropic connection matrix. Let us without proving write down the formulas that replace Eqs. (16) and (17) when the interaction between spins is isotropic.

After removing all the numbers $\mu_{\alpha\beta}$ that due to the symmetry reasons copy the coordinates of the vector $\mathbf{M}$ (see Eq. (12)), in place of (16) we obtain the sequence

$$\{\mu_{1\beta}\}_{\beta=1}^{l+1}, \{\mu_{2\beta}\}_{\beta=2}^{l+1}, \{\mu_{3\beta}\}_{\beta=3}^{l+1}, \ldots, \{\mu_{l\beta}\}_{\beta=l}^{l+1}, \mu_{l+1\,l+1}, \qquad (18)$$

that includes only $(l+1)(l+2)/2$ numbers. Next, when the interaction is isotropic we can rewrite the general requirement (15) as

$$\mu_{11} + 4\left(\sum_{\beta=2}^{l+1}\mu_{1\beta} + \sum_{\alpha=2}^{l+1}\mu_{\alpha\alpha}\right) + 8\sum_{\alpha=2}^{l}\sum_{\beta=\alpha+1}^{l+1}\mu_{\alpha\beta} = 0,$$

and calculate $\mu_{11}$ with the aid of this equation. As a result, we obtain the correct answer: in the sequence (18) the number of independent values is equal to $l(l+3)/2$.

### 4. Three-dimensional Ising model

**1)** We consider a system of spins at the nods of a cubic lattice of the size $L \times L \times L$ ($L = 2l+1$) assuming periodic boundary conditions. Then each spin has $l$ pairs of neighbors that are situated along the three independent coordinate axes. In addition, the spins have neighbors that are not on the same coordinate axes as the given spin. The spins equally interacting with the given spin constitute a coordination sphere.

Let $w(n,m,k)$ be a constant of interaction between spins shifted with respect to each other by a distance $n$ along the first axis, by a distance $m$ along the second axis, and by a distance $k$ along the third axis. When the interaction is anisotropic, there are $(l+1)^3 - 1$ independent interaction constants $\{w(n,m,k)\}_{n,m,k=0}^{l}$, where $-1$ appears since there is no self-interaction and $w(0,0,0) = 0$. In the case of an isotropic interaction, the number of various constants $w(n,m,k)$ is equal to $(l+1)(l+2)(l+3)/6 - 1$.

In paper [3], we showed that the $(L^3 \times L^3)$-dimensional connection matrix $\mathbf{C}_0$ defined by the interaction constants $\{w(n,m,k)\}_{n,k,m=0}^{l}$ is a block-circulant. It eigenvectors $\mathbf{F}_{\alpha\beta\gamma}$ are the Kronecker products of the eigenvectors $\mathbf{f}^{(\alpha)}$ (see Eq. (3)):

$$\mathbf{F}_{\alpha\beta\gamma} = \mathbf{f}^{(\alpha)} \otimes \mathbf{f}^{(\beta)} \otimes \mathbf{f}^{(\gamma)}, \quad \alpha,\beta,\gamma = 1,2,\ldots,L.$$

The vectors $\mathbf{F}_{\alpha\beta\gamma}$ constitute a full set of the eigenvectors of *any* connection matrix of the three-dimensional Ising system and they do not depend on the type of the interaction constants $\{w(n,m,k)\}_{n,k,m=0}^{l}$. Let us write down the eigenvalues of the matrix $\mathbf{C}_0$ obtained in [3]:

$$\mu_{\alpha\beta\gamma} = \sum_{n=0}^{l}\sum_{m=0}^{l}\sum_{k=0}^{l} w(n,m,k) \cdot \lambda_\alpha(n) \cdot \lambda_\beta(m) \cdot \lambda_\gamma(k), \quad \alpha,\beta,\gamma = 1,2,...,L. \tag{19}$$

We use the above-introduced $L^2$-dimensional vectors $\mathbf{\Lambda}(m,k)$ (see Eq. (11)) to generate $L^3$-dimensional vectors $\mathbf{\Lambda}(n,m,k)$ that are the Kronecker products of the vectors $\mathbf{\Lambda}(n)$ and $\mathbf{\Lambda}(m,k)$

$$\mathbf{\Lambda}(n,m,k) = \mathbf{\Lambda}(n) \otimes \mathbf{\Lambda}(m,k) = \begin{pmatrix} \lambda_1(n)\mathbf{\Lambda}(m,k) \\ \lambda_2(n)\mathbf{\Lambda}(m,k) \\ \vdots \\ \lambda_{l+1}(n)\mathbf{\Lambda}(m,k) \\ \lambda_{l+2}(n)\mathbf{\Lambda}(m,k) \\ \vdots \\ \lambda_L(n)\mathbf{\Lambda}(m,k) \end{pmatrix}, \quad n,m,k = 0,1,...,l. \tag{20}$$

The vectors $\mathbf{\Lambda}(n,m,k)$ are mutually orthogonal.

Let us define an $L^3$-dimensional vector $\mathbf{M}$ whose coordinates are the eigenvalues (19):

$$\mathbf{M} = (\mu_{111}..\mu_{1LL}, \mu_{211}..\mu_{2LL}, ..., \mu_{l+1\,11}..\mu_{l+1\,LL}, \mu_{l+2\,11}..\mu_{l+2\,LL}, ..., \mu_{L11}..\mu_{LLL})^+. \tag{21}$$

Then we can rewrite the set of equations (19) in the vector form:

$$\mathbf{M} = \sum_{n=0}^{l}\sum_{m=0}^{l}\sum_{k=0}^{l} w(n,m,k) \cdot \mathbf{\Lambda}(n,m,k) = w(0,0,1) \cdot \mathbf{\Lambda}(0,0,1) + ... + w(0,l,l) \cdot \mathbf{\Lambda}(0,l,l) + \\ + w(1,0,0) \cdot \mathbf{\Lambda}(1,0,0) + ... + w(1,l,l) \cdot \mathbf{\Lambda}(1,l,l) + .... + w(l,0,0) \cdot \mathbf{\Lambda}(l,0,0) + .... + w(l,l,l) \cdot \mathbf{\Lambda}(l,l,l). \tag{22}$$

The equation (22) allows us to solve the inverse problem and calculate the interaction constants $w(n,m,k)$ that define the given set of the eigenvalues $\{\tilde{\mu}_{\alpha\beta\gamma}\}_{\alpha,\beta,\gamma=1}^{L}$ of the connection matrix. Indeed, let us transform this "experimental" spectrum $\{\tilde{\mu}_{\alpha\beta\gamma}\}_{\alpha,\beta,\gamma=1}^{L}$ into an $L^3$-dimensional column-vector $\tilde{\mathbf{M}}$ of the form (21) and use the mutual orthogonality of the vectors $\mathbf{\Lambda}(n,m,k)$. Then we obtain the required interaction constants as the scalar products of the $L^3$-dimensional vectors:

$$w(n,m,k) = \frac{\left(\tilde{\mathbf{M}}, \mathbf{\Lambda}(n,m,k)\right)}{\|\mathbf{\Lambda}(n,m,k)\|}, \quad n,m,k = 0,1,...l. \tag{23}$$

This formula solves the problem of restoring of the interaction constants corresponding to the given spectrum.

**2)** Not any sequence of the numbers $\{\mu_{\alpha\beta\gamma}\}_{\alpha,\beta,\gamma=1}^{L}$ can represent the spectrum of a three-dimensional Ising connection matrix. To start with, the equality

$$\sum_{\alpha=1}^{L}\sum_{\beta=1}^{L}\sum_{\gamma=1}^{L}\mu_{\alpha\beta\gamma} = 0 \tag{24}$$

has to be hold. As in the two-dimensional problem, the cases of anisotropic and isotropic interactions differ significantly. When the interaction is anisotropic, it is easy to list the values $\mu_{\alpha\beta\gamma}$ where we exclude the numbers repeated due the symmetry reasons. This list contains $(l+1)^3$ values

$$\{\mu_{\alpha\beta\gamma}\}_{\alpha,\beta,\gamma=1}^{l+1}.$$

(compare with Eq. (16)). Because of Eq. (24), independent values in this list are one less. For example, we can express $\mu_{111}$ in terms of the other independent values:

$$\mu_{111} = -2\left(\sum_{\beta=2}^{l+1}\mu_{1\beta 1} + \sum_{\gamma=2}^{l+1}\mu_{11\gamma} + \sum_{\alpha=2}^{l+1}\mu_{\alpha 11}\right) - 4\left(\sum_{\beta,\gamma=2}^{l+1}\mu_{1\beta\gamma} + \sum_{\alpha,\gamma=2}^{l+1}\mu_{\alpha 1\gamma} + \sum_{\alpha,\beta=2}^{l+1}\mu_{\alpha\beta 1}\right) - 8\sum_{\alpha,\beta,\gamma=2}^{l+1}\mu_{\alpha\beta\gamma}. \tag{25}$$

The symmetry reasons allow us to restore all the other numbers $\mu_{\alpha\beta\gamma}$.

Consequently, the number of independent values $\mu_{\alpha\beta\gamma}$ equals exactly to the number of the basic, так vectors $\mathbf{\Lambda}(n,m,k)$ (see Eq. (20)), which enter the sum (22) with nonzero coefficients.

When the interaction is isotropic, due to symmetry restrictions only $(l+1)(l+2)(l+3)/6$ values $\mu_{\alpha\beta\gamma}$ may be independent. They are

$$\{\mu_{1\beta\gamma}\}_{\beta=2,\gamma=\beta+1}^{l+1}, \{\mu_{2\beta\gamma}\}_{\beta=3,\gamma=\beta+1}^{l+1},\ldots, \mu_{l+1l+1l+1}.$$

In addition, because of Eq. (24) this number is less by one. The same as we have done previously (see Eq. (25)), we can define, for example, $\mu_{111}$. Then using the remaining independent values with the aid of the symmetry reasons we restore all the other numbers $\mu_{\alpha\beta\gamma}$. Thus, in the given "experimental" set of eigenvalues there must be $(l+1)(l+2)(l+3)/6-1$ independent values and this number exactly matches the number of various coefficients $w(n,m,k)$ in the expansion (22).

## 5. Discussion and conclusions

Connection matrices define the most important characteristics of Ising systems - such as the energies of the states and their distribution, the free energy, and all the macroscopic properties defined by the free energy. All these functions are crucially dependent on the connection matrix whose main characteristics are its eigenvalues and eigenvectors. In papers [1-3], we obtained the expressions for the eigenvalues of the Ising connection matrix $\mathbf{A} = (A_{ij})_{i,j=1}^N$ with an arbitrary long-range interaction in terms of its matrix elements. In the present paper, we solve the inverse problem: we suppose that we know the matrix spectrum and we have to determine the interaction constants providing this spectrum.

We would like to note that the statement of the problem itself is not obvious. The point is that usually to calculate matrix elements of a matrix we have to know not only the eigenvalues but also all its eigenvectors. Indeed, let $\{\lambda_\alpha\}$ and $\{\mathbf{f}^{(\alpha)} = (f_1^{(\alpha)}, f_2^{(\alpha)},\ldots)^+\}_\alpha$ be eigenvalues and eigenvectors of a symmetric matrix $\mathbf{A} = (A_{ij})$, respectively. Then its matrix elements are [5]:

$$A_{ij} = \sum_{\alpha=1} \lambda_\alpha f_i^{(\alpha)} f_j^{(\alpha)}, \quad i,j=1,2,\ldots. \tag{26}$$

On the other hand, at the beginning of each Section we recall that all connection matrices of any $d$-dimensional Ising model are circulants and, consequently, all these matrices have the same set of the eigenvectors [6]. In other words, their eigenvectors are known by default. However, our analysis shows that when calculating the matrix elements the internal symmetry of the problem allows us not to use Eq. (26) but much more simpler and convenient formulas (see Eqs. (9), (14), and (23)). In addition, using the symmetry reasons, we obtain the number and positions of independent values in a given sequence that allows it to be the spectrum of some connection matrix. The problem is solved for the $d$-dimensional hypercube lattices and an arbitrary long-range interaction.

The following question arises: may the connection matrix eigenvalues be useful when calculating the partition function

$$Z_N = \sum_{i=1}^{2^N} e^{\beta(\mathbf{A}\mathbf{s}_i,\mathbf{s}_i)} = e^{\beta(\mathbf{A}\mathbf{s}_1,\mathbf{s}_1)} + e^{\beta(\mathbf{A}\mathbf{s}_2,\mathbf{s}_2)} + \ldots + e^{\beta(\mathbf{A}\mathbf{s}_k,\mathbf{s}_k)} + \ldots ? \tag{27}$$

The first thing that comes to mind is to improve the transfer-matrix method for the partition function calculation using the expansion (26) of the matrix $\mathbf{A}$ in terms of the eigenvectors and the eigenvalues we obtained. Apparently, this is a hopeless idea. The basis of the transfer-matrix method is a transition from $N$ spin variables to a larger their number $N+1$. However, an increase of the dimension of the problem leads to a complex restructuring of the eigenvalues and the eigenvectors of the matrix $\mathbf{A}$ in each term of Eq. (27).

On the other hand, the connection matrix eigenvalues themselves may be useful when calculating the partition functions. Indeed, let us expand each exponential in (27) into a formal Taylor series. Now let us rearrange the summands, combining in one sum the terms with the same power $\beta$. We showed that for matrices with zero diagonals the equation

$$Z_N \equiv 2^N + \beta \cdot \sum_{i=1}^{2^N}(\mathbf{A}\mathbf{s}_i,\mathbf{s}_i) + \frac{\beta^2}{2!}\sum_{i=1}^{2^N}(\mathbf{A}\mathbf{s}_i,\mathbf{s}_i)^2 + \frac{\beta^3}{3!}\sum_{i=1}^{2^N}(\mathbf{A}\mathbf{s}_i,\mathbf{s}_i)^3 + \ldots = 2^N\left(1 + \beta\operatorname{Tr}\mathbf{A} + \beta^2\operatorname{Tr}\mathbf{A}^2 + \frac{4}{3}\beta^3\operatorname{Tr}\mathbf{A}^3 + \ldots\right), \tag{28}$$

is true [8]. Here Tr means the trace of the matrix. Since $\operatorname{Tr}\mathbf{A} = 0$ the summand $\beta\operatorname{Tr}\mathbf{A}$ in the right-hand side of Eq. (28) is equal to zero too. Then the partition function is a series in the powers of the inverse temperature $\beta$ with the coefficients $\operatorname{Tr}\mathbf{A}^k = \sum_{i=1}^N \lambda_i^k$ that are the sums of the powers of the connection matrix eigenvalues. In our case, the eigenvalues $\lambda_i$ are determined by the equations (4), (10) or (19). Note, beginning from $k=4$ the expressions for the sums $\sum_{i=1}(\mathbf{As}_i, \mathbf{s}_i)^k$ become more complex including not only $\operatorname{Tr}\mathbf{A}^k$ but also some additional terms.

Our arguments show that the eigenvalues of the Ising connection matrix may be useful when calculating the partition function.

**Acknowledgements**

Funding: This work was supported by Russian Basic Research Foundation gran # 18-07-00500.
We are grateful to Dr. Inna Kaganova for help when preparing this paper.